\documentclass[12pt]{article}

\begin{document}

\begin{center}
{\bf Generalized Weinberg-Tucker-Hammer equations in the
Petiau-Duffin-Kemmer form} \\
\vspace{5mm} S. I. Kruglov\\
%\footnote{E-mail: krouglov@utsc.utoronto.ca}
\vspace{3mm} \textit{University of Toronto at Scarborough,\\
Physical and Environmental Sciences Department, \\
1265 Military Trail, Toronto, Ontario, Canada M1C 1A4} \\

\vspace{5mm}
\end{center}

\begin{abstract}
Massive and massless vector particles, in the framework of
generalized Weinberg-Tucker-Hammer equations, are considered in
the Petiau-Duffin-Kemmer formalism. We obtain solutions of
equations in the form of matrix-dyads. The Lagrangian is defined
in the matrix form and the conserved electric current and the
energy-momentum tensor are obtained. The canonical quantization is
performed and the propagator of massive fields is found in the
formalism considered.
\end{abstract}

\section{Introduction}

The vector fields play very important role in the Standard Model
which unifies weak and electromagnetic interactions of elementary
particles. After the spontaneous breaking symmetry massless vector
fields acquire masses due to the Higgs mechanism. But the
existence of scalar Higgs bosons in nature is questionable and is
verified at the Large Hadronic Collider. Therefore, the
alternative description of massive and massless vector particles
is of great theoretical interest. The second-order Proca
equations, that are used in the theory of vector particles, can be
represented in the form of the first-order Petiau-Duffin-Kemmer
(PDK) relativistic wave equation \cite{Petiau}, \cite{Duffin},
\cite{Kemmer} (see also \cite{Umezawa}). The PDK form of equations
is very convenient for different applications \cite{Krajcik}. We
have used the PDK form of equations in the Dirac-K\"{a}hler theory
\cite{Kruglov}, in the field theory of multi-spin particles
\cite{Kruglov1}, \cite{Kruglov1'}, and in generalized
electrodynamics \cite{Kruglov2}.

In this paper, we consider generalized Weinberg-Tucker-Hammer
equations (WTH) for massive and massless vector fields, introduced
by Dvoeglazov \cite{Dvoeglazov}, in the PDK formalism. We obtain
solutions in the form of projection matrix-dyads, find the
conserved electric current and the energy-momentum tensor, and
perform the canonical quantization in the PDK form.

The paper is organized as follows. In Sec.2, we represent
generalized WTH equations for massive and massless vector fields
in the form of generalized PDK equations. Solutions of the wave
equation for a free fields are obtained in the form of
matrix-dyads in Sec.3. We define the Lagrangian in the matrix form
and find the conserved electric current and the energy-momentum
tensor in Sec.4. The canonical quantization of fields is performed
and the propagator is obtained in the PDK form in Sec.5. We
discuss the results obtained in Sec.6. Some useful matrix
relations are found in Appendix.

The Euclidean metric is used and the system of units $\hbar =c=1$
is explored. Greek and Latin letters run $1,2,3,4$ and $1,2,3$,
respectively, and, we imply the summation in the repeated indexes.

\section{Generalized WTH equations in the PDK formalism}

Let us consider generalized WTH equations for vector fields,
introduced by Dvoeglazov \cite{Dvoeglazov}:
\begin{equation}
\left(\gamma_{\alpha\beta}\partial
_\alpha\partial_\beta+A\partial_\alpha^2-Bm^2\right) \psi(x) =0,
\label{1}
\end{equation}
where, $\gamma_{\alpha\beta}$ are $6\times 6$-matrices given in
\cite{Barut}, $\partial_\mu=(\partial/\partial
x_m,\partial/\partial (it))$, $A$, $B$ are dimensionless
parameters, and $m$ is the mass parameter. The wave function
$\psi=$column$(\textbf{E}+i\textbf{B},\textbf{E}-i\textbf{B})$,
(where $E_a=iF_{a4}$, $B_a=(1/2)\epsilon_{abc}F_{bc}$) realizes
the $(1,0)\oplus(0,1)$ representation of the Lorentz group. Eq.(1)
is the generalization of the Weinberg \cite{Weinberg} and
Tucker-Hammer equations \cite{Hammer}. At $A=0$, $B=1$ one arrives
at the Weinberg equation, and at $A=1$, $B=2$, we have the
Tucker-Hammer equation \cite{Dvoeglazov}. In the component form
Eq.(1) is equivalent to the second-order equation for the
antisymmetric tensor $F_{\alpha\beta}$ \cite{Dvoeglazov}:
\begin{equation}
\partial_\alpha\partial_\mu F_{\mu\beta}-\partial_\beta\partial_\mu F_{\mu\alpha}
+\frac{A-1}{2}\partial_\mu^2F_{\alpha\beta}-\frac{B}{2}m^2F_{\alpha\beta}=0,
\label{2}
\end{equation}
We study Eq.(2) in the PDK form as for the massive case, $m\neq
0$, as well as for the massless case, $m=0$.

\subsection{Massive vector fields}

We introduce the $10$-component wave function
\begin{equation}
\Psi (x)=\left\{ \psi _A(x)\right\} =\left(
\begin{array}{c}
\psi_\mu (x)\\
\psi_{[\mu\nu]}(x)\\
\end{array}
\right) =\left(
\begin{array}{c}
\frac{1}{m}\partial_\nu F_{\mu\nu} (x)\\
F_{\mu\nu}(x)\\
\end{array}
\right) \hspace{0.3in}(A=\mu , [\mu\nu]).\label{3}
\end{equation}
With the help of the elements of the entire matrix algebra
$\varepsilon ^{A,B}$, with products and matrix elements
\begin{equation}
\left( \varepsilon ^{M,N}\right) _{AB}=\delta _{MA}\delta _{NB},
\hspace{0.5in}\varepsilon ^{M,A}\varepsilon ^{B,N}=\delta
_{AB}\varepsilon ^{M,N}, \label{4}
\end{equation}
where $A,B,M,N=\mu,[\mu\nu]$, Eq.(2) can be represented in the
form
\[
\biggl[\partial _\mu \left(\varepsilon ^{\nu,[\nu\mu]}+
\varepsilon ^{[\nu\mu],\nu}\right) + m\varepsilon ^{\mu,\mu}
\]
\vspace{-6mm}
\begin{equation} \label{5}
\end{equation}
\vspace{-6mm}
\[
+\left(\frac{1-A}{2m}\partial_\mu^2+\frac{B}{2}m\right)\frac{1}{2}\varepsilon
^{[\nu\mu],[\nu\mu]}\biggr] _{AB}\Psi _B(x)=0 .
\]
We imply a summation over all repeated indices. Defining the
$10\times 10$ matrices
\begin{equation}
\beta_\mu=\varepsilon ^{\nu,[\nu\mu]}+ \varepsilon
^{[\nu\mu],\nu},~~~\overline{P}=\varepsilon ^{\mu,\mu},~~~
P=\frac{1}{2}\varepsilon ^{[\nu\mu],[\nu\mu]}, \label{6}
\end{equation}
Eq.(5) takes the PDK matrix form
\begin{equation}
\left[ \beta _\mu \partial _\mu + m\overline{P}+
P\left(\frac{1-A}{2m}\partial_\mu^2+\frac{B}{2}m\right)\right]
\Psi (x)=0 , \label{7}
\end{equation}
where the matrices $\beta_\mu$ are Hermitian matrices,
$\beta_\mu^+=\beta_\mu$. Eq.(7) is the generalized PDK equation.
The projection operator $\overline{P}= \overline{P}^+$ extracts
the four-dimensional vector subspace ($\psi_\mu$) of the wave
function $\Psi$, and the projection operator $P=P^+$ extracts the
six-dimensional tensor subspace corresponding to the
$\psi_{[\mu\nu]}$. The matrices $\beta _\mu$ obey the PDK algebra
\begin{equation}
\beta _\mu \beta _\nu \beta _\alpha +\beta _\alpha \beta _\nu
\beta _\mu =\delta _{\mu \nu }\beta _\alpha+\delta _{\alpha \nu
}\beta _\mu , \label{8}
\end{equation}
and matrices $\overline{P}$, $P$ are projection matrices
\[
\overline{P}^2=\overline{P},~~P^2=P,~~\overline{P}+P=1,~~
\overline{P}P=P\overline{P}=0,
\]
\vspace{-8mm}
\begin{equation}
\label{9}
\end{equation}
\vspace{-8mm}
\[
\beta_\mu\overline{P}+\overline{P}\beta_\mu=\beta_\mu,~~ \beta_\mu
P+P\beta_\mu=\beta_\mu.
\]

The Lorentz group generators in the $10$-dimension representation
space are  given by \cite{Kruglov1}, \cite{Kruglov1'}
\begin{equation}
J_{\mu\nu}=\beta_\mu\beta_\nu-\beta_\nu\beta_\mu=
\varepsilon^{[\lambda\mu],[\lambda\nu]}+\varepsilon^{\mu,\nu}-
\varepsilon^{[\lambda\nu],[\lambda\mu]}-\varepsilon^{\nu,\mu},
 \label{10}
\end{equation}
and obey the commutation relations
\[
\left[J_{\rho\sigma},J_{\mu\nu}\right]=\delta_{\sigma\mu}J_{\rho\nu}+
\delta_{\rho\nu}J_{\sigma\mu}-\delta_{\rho\mu}J_{\sigma\nu}-\delta_{\sigma\nu}J_{\rho\mu},
\]
\vspace{-8mm}
\begin{equation}
\label{11}
\end{equation}
\vspace{-8mm}
\[
\left[\beta_{\lambda},J_{\mu\nu}\right]=\delta_{\lambda\mu}\beta_\nu
-\delta_{\lambda\nu}\beta_\mu,~~\left[\overline{P},J_{\mu\nu}\right]=
0,~~\left[P,J_{\mu\nu}\right]= 0 .
\]
Eq.(7) is form-invariant because of Eq.(11). The Lorentz-invariant
is $\overline{\Psi }\Psi =\Psi ^{+}\eta \Psi$, where $\Psi ^{+}$
is the Hermitian-conjugated wave function, and the Hermitianizing
matrix $\eta $ is
\begin{equation}
\eta =\varepsilon ^{m,m}-\varepsilon ^{4,4} +\varepsilon
^{[m4],[m4]}-\frac{1}{2}\varepsilon ^{[mn],[mn]}. \label{12}
\end{equation}
The $\eta $ is the Hermitian matrix, $\eta^+ =\eta $, and obeys
the relations: $\eta \beta _m=-\beta _m\eta$ (m=1,2,3), $\eta
\beta _4=\beta _4\eta$. With the help of these relations, one
finds the ``conjugated" equation
\begin{equation}
\overline{\Psi}(x)\left[ \beta _\mu \overleftarrow{\partial}_\mu -
m\overline{P}-
\left(\overleftarrow{\partial_\mu^2}\frac{1-A}{2m}+\frac{B}{2}m\right)P\right]
=0 , \label{13}
\end{equation}
where $\overline{\Psi }=(\psi_\mu^*,-\psi_{[\mu\nu]}^*)$ and the
complex conjugation $*$ does not act on the metric imaginary unit
$i$ of fourth components of the wave function,
$\psi_\mu^*=(\psi_m^*,i\psi_0^*)$, and so on.

\subsection{Massless vector fields}

To get the massless WTH equation, we put $m=0$ in Eq.(2) and
arrive at
\begin{equation}
\partial_\alpha\partial_\mu F_{\mu\beta}-\partial_\beta\partial_\mu F_{\mu\alpha}
+\frac{A-1}{2}\partial_\mu^2F_{\alpha\beta}=0. \label{14}
\end{equation}
In the same manner as for the massive case, we introduce the
$10$-component wave function
\begin{equation}
\Psi (x)=\left\{ \psi _A(x)\right\} =\left(
\begin{array}{c}
\psi_\mu (x)\\
\psi_{[\mu\nu]}(x)\\
\end{array}
\right) =\left(
\begin{array}{c}
\frac{1}{\kappa}\partial_\nu F_{\mu\nu} (x)\\
F_{\mu\nu}(x)\\
\end{array}
\right),\label{15}
\end{equation}
where $\kappa$ is the new parameter with the dimension of the
mass. We need the dimensional parameter to have the components of
the wave function $\Psi(x)$ with the same dimension. Physical
values do not depend on this parameter, and $\kappa$ is absorbed
by the renormalization of fields \cite{Kruglov2}. As a result,
Eq.(14) takes the form similar to Eq.(7) for massive fields
\begin{equation}
\left[ \beta _\mu \partial _\mu + \kappa\overline{P}+
P\frac{C}{\kappa}\partial_\mu^2\right] \Psi (x)=0 , \label{16}
\end{equation}
where $C$ is an arbitrary parameter ($C=(A-1)/2$). Thus, the
equation of motion and other equations for the massless fields can
be obtained from the massive case by the replacements
$m\rightarrow\kappa$, $B=0$. Therefore, in the further
consideration, we write down only the expressions for the massive
case implying that they are valid also for the massless fields
after the replacements $m\rightarrow\kappa$, $B=0$. Eq.(14), (16)
represent the modified Maxwell equations.

\section{Solutions of the matrix equation}

In the momentum space, for the positive ($+p$) and negative ($-p$)
energies, Eq.(7) reads
\begin{equation}
\left( \pm i\widehat{p} + m\overline{P}+ \lambda P\right) \Psi
(\pm p)=0 , \label{17}
\end{equation}
where $\widehat{p}=\beta _\mu p _\mu$, the four-momentum is
$p_\mu=(\textbf{p},ip_0)$, and
\begin{equation}
\lambda=\frac{A-1}{2m}p^2+\frac{B}{2}m, \label{18}
\end{equation}
$p^2= \textbf{p}^2-p_0^2$. The matrix of Eq.(17)
\begin{equation}
\Lambda_\pm = \pm i\widehat{p} + m\overline{P}+ \lambda P,
\label{19}
\end{equation}
obeys the minimal polynomial (see Eq.(A3) in Appendix):
\begin{equation}
\left(\Lambda_\pm-m\right)\left(\Lambda_\pm-\lambda \right)
\left[\left(\Lambda_\pm-m\right)\left(\Lambda_\pm-\lambda
\right)+p^2 \right]=0. \label{20}
\end{equation}
There exist non-trivial solutions of Eq.(17) if $\det \Lambda_\pm
=0$ or the eigenvalue of the $\Lambda_\pm$ is equal to zero. If
$\lambda\neq 0$, this requirement leads to the dispersion relation
\begin{equation}
p^2+\lambda m=0 . \label{21}
\end{equation}
It follows from Eq.(20) that there are also other eigenvalues of
the matrix $\Lambda_\pm$: $m$ and $\lambda$. Taking into account
Eq.(18), one obtains from Eq.(21)
\begin{equation}
p^2+\frac{m^2B}{A+1}=0 . \label{22}
\end{equation}
Thus, the mass of the vector field is
\begin{equation}
M=m\sqrt{\frac{B}{A+1}} . \label{23}
\end{equation}
Another case which leads to $\det \Lambda_\pm =0$, is $\lambda=0$.
Then Eq.(18) results at $\lambda=0$ to the physical mass
\begin{equation}
M'=m\sqrt{\frac{B}{A-1}} . \label{24}
\end{equation}
The formulas (23), (24) are in the agreement with results obtained
by Dvoeglazov \cite{Dvoeglazov}. From Eq.(23), we find the
restriction on the parameters $A$ and $B$: $B/(A+1)\geq 0$, and in
the case of Eq.(24), $B/(A-1)\geq 0$. It follows from Eq.(23) that
for Weinberg \cite{Weinberg} ($A=0$, $B=1$) and Tucker-Hammer
\cite{Hammer} ($A=1$, $B=2$) equations, we have the same physical
mass of vector particles, $M=m$. The case (24) is not realized for
Tucker-Hammer equations \cite{Hammer} ($A=1$, $B=2$, $\lambda\neq
0$). Therefore, we imply that $\lambda\neq 0$ and the physical
masses of vector particles are given by Eq.(23). On-shell, when
Eq.(22) is valid, and $m\neq\lambda$ the minimal polynomial
equation (20) reduces to
\begin{equation}
\Lambda_\pm\left(\Lambda_\pm-\lambda-m \right)
\left(\Lambda_\pm-m\right)\left(\Lambda_\pm-\lambda \right)=0.
\label{25}
\end{equation}
Eq.(25) allows us to obtain solutions of Eq.(17),
$\Lambda\Psi(p)=0$, in the form of the projection matrix
\cite{Fedorov}. Indeed, the matrix
\[
\Pi_\pm=N\left(\Lambda_\pm-\lambda-m \right)
\left(\Lambda_\pm-m\right)\left(\Lambda_\pm-\lambda \right)
\]
\vspace{-8mm}
\begin{equation}
\label{26}
\end{equation}
\vspace{-8mm}
\[
=\mp Ni\widehat{p}\left(\pm im\widehat{p}P\pm
i\lambda\widehat{p}\overline{P} -\lambda m\right),
\]
where $N$ is the normalization constant, obeys the equation
$\Lambda_\pm\Pi_\pm=0$. It means that every column of the matrix
$\Pi_\pm$ is the solution to Eq.(17). The requirement that the
$\Pi_\pm$ is the projection matrix
\begin{equation}
\Pi_\pm^2=\Pi_\pm, \label{27}
\end{equation}
gives the value
\begin{equation}
N=-\frac{1}{\lambda m(\lambda+m)} . \label{28}
\end{equation}
Eq.(28) can be verified with the help of the minimal Eq.(25). It
should be noted that Eq.(25) represents the minimal polynomial for
the non-degenerate case $m\neq\lambda$ (see Eq.(A3) in Appendix).
But the case $m=\lambda$ is realized as for the Weinberg equation
($A=0$, $B=1$) as well as for the Tucker-Hammer equation ($A=1$,
$B=2$). This follows from Eq.(18),(23). For this special case,
$m=\lambda $, one should explore the minimal polynomial equation
on-shell (see Eq.(A4) in Appendix)
\begin{equation}
\Lambda_\pm\left(\Lambda_\pm-m\right)\left(\Lambda_\pm-2m\right)=0.
\label{29}
\end{equation}
which corresponds to the PDK equation for spin-1. Indeed, if
$m=\lambda$, Eq.(17) becomes
\begin{equation}
\left( \pm i\widehat{p} + m \right) \Psi (\pm p)=0 , \label{30}
\end{equation}
as $\overline{P}+P=1$. But Eq.(30) is the PDK equation for vector
fields and the solution in the form of the matrix-dyad to this
equation is well known \cite{Fedorov}. We find the projection
operator from Eq.(29)
\begin{equation}
\Pi^{PDK}_\pm=\frac{1}{2m^2}\left(\Lambda_\pm^2-3m
\Lambda_\pm+2m^2\right) =\frac{\pm i\widehat{p}\left(\pm
i\widehat{p}-m\right)}{2m^2}. \label{31}
\end{equation}
Eq.(26) converts to Eq.(31) at $m=\lambda$.

The projection operators extracting spin projections $\pm 1$ and
$0$ are given by \cite{Fedorov},
\begin{equation}
S_{(\pm 1)}=\frac 12\sigma _p\left( \sigma _p\pm 1\right)
,\hspace{0.5in}S_{(0)}=1-\sigma _p^2 .\label{32}
\end{equation}
where the spin operator is
\begin{equation}
\sigma _p=-\frac i{2\mid \mathbf{p}\mid }\epsilon
_{abc}p_aJ_{bc}=-\frac i{\mid \mathbf{p}\mid }\epsilon
_{abc}p_a\beta _b\beta _c \label{33}
\end{equation}
and commutes with the matrix of Eq.(17) $\Lambda_\pm$:
$[\Lambda_\pm,\sigma _p]=0$. The projection operators satisfy the
relations: $S_{(\pm 1)}^2=S_{(\pm 1)}$, $S _{(\pm 1)}S_{(0)}=0$,
$S_{(0)}^2=S_{(0)}$. The operators (32) commute with the mass
projection operator (26) and, as a result, we obtain projection
operators extracting solutions to Eq.(17) for spin projections
$\pm 1$, $0$ (for states of particles with the mass $m$) in the
form of matrix-dyads
\begin{equation}
\Pi_\pm S_{(\pm 1)}=\Psi_{\pm 1}\cdot \overline{\Psi}_{\pm 1}
,~~~~\Pi_\pm S_{(0)}=\Psi_{0}\cdot \overline{\Psi}_{0}. \label{34}
\end{equation}
The matrix elements of the matrix-dyad is $(\Psi\cdot
\overline{\Psi})_{AB}= \Psi_A \overline{\Psi}_B $. For the case of
the Weinberg and Tucker-Hammer equations, one has to make the
replacement in Eq.(34) $\Pi_\pm\rightarrow \Pi_\pm^{PDK}$. Eq.(34)
can be used for different electrodynamics calculations of
processes with vector particles in the framework of generalized
WTH model.

For the case of massless WTH equation, the parameter on-shell is
$\lambda =0$ ($p^2=0)$. Then the operator of the equation becomes
\begin{equation}
\Lambda_\pm= i\widehat{p} + \kappa\overline{P}, \label{35}
\end{equation}
and obeys the minimal matrix equation on-shell (see Eq.(A6) in
Appendix) as follows:
\begin{equation}
\Lambda_\pm^2\left(\Lambda_\pm-\kappa\right)^2 =0. \label{36}
\end{equation}
As pointed in \cite{Moroz}, in this case there are difficulties to
obtain the solutions in the form of projection operators because
the zero eigenvalue of the operator $\Lambda_\pm$ is degenerated.
In the case of Maxwell equations such difficulty was overcome in
\cite{Kruglov2} by using the general gauge and adding a scalar
field.

\section{Lagrangian and conserved currents}

The Lagrangian of generalized WTH model in the PDK formalism is
given by
\[
\mathcal{L}=-\frac{1}{2}\overline{\Psi }(x) \beta _\mu
\left(\partial _\mu-\overleftarrow{\partial}_\mu\right)\Psi (x) -
m\overline{\Psi }(x)\left(\overline{P} + \frac{B}{2} P\right)\Psi
(x)
\]
\vspace{-8mm}
\begin{equation}
\label{37}
\end{equation}
\vspace{-8mm}
\[
+ \frac{1-A}{2m}\left(\partial_\mu\overline{\Psi
}(x)\right)P\left(\partial_\mu\Psi (x)\right) .
\]
One may construct another Lagrangian by adding the four-divergence
to Eq.(37) that does not change equations of motion. It can be
verified that Euler-Lagrange equations following from Eq.(37) give
the equations of motion (7),(13). We obtain also the Lagrangian
(37) in terms of fields $\psi_A$
\[
{\cal L}=\frac{1}{2}\biggl[\psi_{[\rho\mu]}^*\partial
_\mu\psi_\rho-\psi_\rho^*\partial _\mu\psi_{[\rho\mu]} - m
\psi_\mu^*\psi_\mu
\]
\vspace{-8mm}
\begin{equation}
\label{38}
\end{equation}
\vspace{-8mm}
\[
+\frac{mB}{2}\psi_{[\rho\mu]}^*\psi_{[\rho\mu]}-\frac{1-A}{2m}
\left(\partial_\nu\psi_{[\rho\mu]}^*\right)\left(\partial_\nu\psi_{[\rho\mu]}\right)\biggr]
+ c.c. ,
\]
where the $c.c.$ is the complex conjugated expression. The
electric current density is given by \cite{Ahieser}
\begin{equation}
j_\mu (x)=i\left( \overline{\Psi }(x)
\frac{\partial\mathcal{L}}{\partial\left(\partial_\mu
\overline{\Psi} (x)\right)}-
\frac{\partial\mathcal{L}}{\partial\left(\partial_\mu \Psi
(x)\right)}\Psi (x)\right) . \label{39}
\end{equation}
Replacing Eq.(37) into Eq.(39), we obtain the electric current
density
\begin{equation}
j_\mu (x)=i\overline{\Psi}(x)\beta_\mu \Psi(x) + \frac{1-A}{2m}
\left[\overline{\Psi}(x)P\partial_\mu
\Psi(x)-\left(\partial_\mu\overline{\Psi}(x)\right)P\Psi(x)\right].
\label{40}
\end{equation}
One may check with the help of equations of motion (7),(13) that
the electric current is conserved, $\partial_\mu j_\mu (x)=0$. The
second term in Eq.(40) vanishes at $A=1$, corresponding to the
Tucker-Hammer equation, and $j_\mu (x)$ takes the form similar to
the Dirac theory. Using the general expression for the canonical
energy-momentum tensor \cite {Ahieser}
\begin{equation}
T_{\mu\nu}=\frac{\partial\mathcal{L}}{\partial\left(\partial_\mu
\Psi (x)\right)}\partial_\nu \Psi (x)+\partial_\nu \overline{\Psi
}(x) \frac{\partial\mathcal{L}}{\partial\left(\partial_\mu
\overline{\Psi} (x)\right)}-\delta_{\mu\nu} \mathcal{L}
,\label{41}
\end{equation}
we obtain from the Lagrangian (37)
\[
 T_{\mu\nu}=\frac{1}{2}\left(\partial_\nu \overline{\Psi}
(x)\right)\beta_\mu \Psi (x)-\frac{1}{2} \overline{\Psi}
(x)\beta_\mu \partial_\nu\Psi (x)
\]
\vspace{-8mm}
\begin{equation}
\label{42}
\end{equation}
\vspace{-8mm}
\[
+ \frac{1-A}{2m}\left[\left(\partial_\mu\partial_\nu\overline{\Psi
}(x)\right)P\Psi (x)+\overline{\Psi
}(x)P\partial_\mu\partial_\nu\Psi (x)\right]-\delta_{\mu\nu}
\mathcal{L} .
\]
One may verify that the canonical energy-momentum tensor is
conserved tensor, $\partial_\mu T_{\mu\nu}=0$. At $A=1$
Eq.(37),(42) are simplified. It should be noted that the
Lagrangian $\mathcal{L}$, Eq.(37), does not vanish if $A\neq 1$
for fields satisfying the equations of motion. It is easy to write
down expressions for $j_\mu$ and $T_{\mu\nu}$ in the component
form exploring Eq.(3),(4),(6).

\section{The canonical quantization}

Solutions to Eq.(7) with definite energy and momentum in the form
of plane waves are given by
\begin{equation}
\Psi_s^{(\pm)}(x)=\sqrt{\frac{m+\lambda}{2p_0 V}}\Psi_s(\pm
p)\exp(\pm ipx) , \label{43}
\end{equation}
where $V$ is the normalization volume, $p^2=
\textbf{p}^2-p_0^2=-m^2$, and $s$ is the spin index ($s=\pm 1,0$).
The 10-dimensional function $\Psi_s(\pm p)$ obeys Eq.(17). We use
the normalization conditions
\begin{equation}
\int_V \overline{\Psi}^{(\pm)}_{s}(x)\beta_4 \Psi^{(\pm)}_{s'
}(x)d^3 x=\pm\delta_{ss'} ,~~~~\int_V
\overline{\Psi}^{(\pm)}_{s}(x)\beta_4 \Psi^{(\mp)}_{s' }(x)d^3 x=0
, \label{44}
\end{equation}
where $\overline{\Psi}^{(\pm)}_{s}(x)=\left(\Psi^{(\pm)}_{s
}(x)\right)^+ \eta$. Eq.(44) correspond to the normalization on
the charge. In the second quantized theory, the field operators
are represented by
\[
\Psi(x)=\sum_{p,s}\left[a_{p,s}\Psi^{(+)}_{s}(x) +
b^+_{p,s}\Psi^{(-)}_{s}(x)\right] ,
\]
\vspace{-7mm}
\begin{equation} \label{45}
\end{equation}
\vspace{-7mm}
\[
\overline{\Psi}(x)=\sum_{p,s}\left[a^+_{p,s}
\overline{\Psi}^{(+)}_{s}(x)+ b_{p,s}
\overline{\Psi}^{(-)}_{s}(x)\right] ,
\]
where the positive and negative parts of the wave function are
given by Eq.(43). The creation and annihilation operators of
particles, $a^+_{p,s}$, $a_{p,s}$, and antiparticles, $b^+_{p,s}$,
$b_{p,s}$, obey the commutation relations as follows:
\[
[a_{p,s},a^+_{p',s'}]=\delta_{ss'} \delta_{pp'} ,~~~
[a_{p,s},a_{p',s'}]=[a^+_{p,s},a^+_{p',s'}]=0,
\]
\begin{equation}
[b_{p,s},b^+_{p',s'}]=\delta_{ss'} \delta_{pp'} ,~~~
[b_{p,s},b_{p',s'}]=[b^+_{p,s},b^+_{p',s'}]=0, \label{46}
\end{equation}
\[
[a_{p,s},b_{p',s'}]=[a_{p,s},b^+_{p',s'}]=
[a^+_{p,s},b_{p',s'}]=[a^+_{p,s},b^+_{p',s'}]=0.
\]
We obtain the commutation relations for different times from
Eq.(43)-(46)
\[
\left[\Psi_M(x),\Psi_N(x')]=[\overline{\Psi}_{ M}(x),
\overline{\Psi}_{ N}(x')\right] =0, ~~\left[\Psi_{
M}(x),\overline{\Psi}_{N}(x')\right]=N_{ MN}(x,x'),
\]
\[
N_{MN}(x,x')=N^+_{MN}(x,x')-N^-_{ MN}(x,x'),
\]
\vspace{-7mm}
\begin{equation} \label{47}
\end{equation}
\vspace{-7mm}
\[
N^+_{MN}(x,x')=\sum_{p,s}\left(\Psi^{(+)}_{s}(x)\right)_M
\left(\overline{\Psi}^{(+)}_{s}(x')\right)_N ,
\]
\[
N^-_{MN}(x,x')=\sum_{p,s}\left(\Psi^{(-)}_{s}(x)\right)_M
\left(\overline{\Psi}^{(-)}_{s}(x')\right)_N .
\]
One finds from Eq.(43),(47)
\begin{equation}
N^\pm_{MN}(x,x')=\sum_{p,s}\frac{m+\lambda}{2p_0
V}\left(\Psi_{s}(\pm p)\right)_M\left(\overline{\Psi}_{s}(\pm
p)\right)_N\exp [\pm ip(x-x')] .
 \label{48}
\end{equation}
It follows from Eq.(32) that the equation
\begin{equation}
S_{(+ 1)}+ S_{(- 1)}+S_{(0)}=1,\label{49}
\end{equation}
holds, and as a result, we obtain from Eq.(34)
\begin{equation}
\sum_s\left(\Psi_{s}(\pm
p)\right)\cdot\left(\overline{\Psi}_{s}(\pm p)\right) =
\Pi_{\pm}=\frac{\pm i\widehat{p}\left(\pm im\widehat{p}P\pm
i\lambda\widehat{p}\overline{P} -\lambda m\right)}{\lambda
m\left(m+\lambda\right)}. \label{50}
\end{equation}
Taking into account Eq.(50), one finds from Eq.(48):
\[
N^\pm_{MN}(x,x')=\sum_{p} \frac{1}{2p_0 V} \left(\frac{\pm
i\widehat{p}\left(\pm im\widehat{p}P\pm
i\lambda\widehat{p}\overline{P} -\lambda m\right)}{\lambda
m}\right)_{MN} \exp [\pm ip(x-x')]
\]
\vspace{-6mm}
\begin{equation} \label{51}
\end{equation}
\vspace{-6mm}
\[
= \left( \frac{\beta_\mu\partial_\mu\left(m\beta_\mu P\partial_\mu
+ \lambda\beta_\mu \overline{P}\partial_\mu -\lambda
m\right)}{\lambda m} \right)_{MN} \Delta_{\pm}(x-x'),
\]
where the singular functions are given by \cite{Ahieser}
\begin{equation}
\Delta_+(x)=\sum_{p}\frac{1}{2p_0V}\exp
(ipx),~~~~\Delta_-(x)=\sum_{p}\frac{1}{2p_0V}\exp (-ipx).
\label{52}
\end{equation}
It follows from Eq.(18) that the $\lambda$ (at $ A\neq 1$) is the
operator because in the configuration space $p^2=-\partial_\mu^2$.
With the help of the function \cite{Ahieser}
\begin{equation}
\Delta_0 (x)=i\left(\Delta_+(x)-\Delta_-(x)\right), \label{53}
\end{equation}
from Eq.(47),(51),(53), we obtain
\begin{equation}
N_{MN}(x,x')=-i\left(\frac{\beta_\mu\partial_\mu\left(m\beta_\mu
P\partial_\mu + \lambda\beta_\mu \overline{P}\partial_\mu -\lambda
m\right)}{\lambda m}\right)_{MN} \Delta_{0}(x-x').
 \label{54}
\end{equation}
It should be noted that the function $\Delta_{0}(x)$ vanishes when
$x^2= \textbf{x}^2-t^2>0$ \cite{Ahieser}. Using Eq.(8), one finds
the relation
\[
 \left(\beta_\mu\partial_\mu+ m
\overline{P}+\lambda P
\right)\frac{\beta_\mu\partial_\mu\left(m\beta_\mu P\partial_\mu +
\lambda\beta_\mu \overline{P}\partial_\mu -\lambda
m\right)}{\lambda m}
\]
\vspace{-7mm}
\begin{equation} \label{55}
\end{equation}
\vspace{-7mm}
\[
= \frac{\left( m \overline{P}+\lambda
P\right)\beta_\mu\partial_\mu }{\lambda
m}\left(\partial_\mu^2-\lambda m\right).
 \]
Taking into account the equation \cite{Ahieser}
$\left(\partial_\mu^2-\lambda m\right)\Delta_\pm (x)=0$, and
Eq.(51), from Eq.(55), we arrive at
\begin{equation}
\left(\beta_\mu\partial_\mu + m \overline{P}+\lambda
P\right)N^\pm(x,x')=0 .
 \label{56}
\end{equation}
For bosonic fields the chronological product of two operators is
defined as \cite{Ahieser}
\[
T\Psi_{M}(x_1)\overline{\Psi}_{ N}(x_2)= \left \{
\begin{array}{c}
\Psi_{M}(x_1)\overline{\Psi}_{N}(x_2),~~~~t_1>t_2,\\
\overline{\Psi}_{N}(x_2)\Psi_{M}(x_1),~~~~t_2>t_1\\
\end{array}
\right \}.
\]
Then the vacuum expectation of chronological pairing of operators
(the propagator) becomes
\[
\langle T\Psi_{M}(x)\overline{\Psi}_{ N}(y)\rangle_0=N^c_{
MN}(x-y)
\]
\vspace{-8mm}
\begin{equation} \label{57}
\end{equation}
\vspace{-8mm}
\[
=\theta\left(x_0 -y_0\right)N^+_{MN}(x-y)+\theta\left(y_0
-x_0\right)N^-_{MN}(x-y) ,
\]
where $\theta(x)$ being the theta-function. We obtain from
Eq.(57):
\begin{equation}
\langle T\Psi(x)\cdot\overline{\Psi}(y)\rangle_0
=\frac{\beta_\mu\partial_\mu\left(m\beta_\mu P\partial_\mu +
\lambda\beta_\mu \overline{P}\partial_\mu -\lambda
m\right)}{\lambda m}\Delta_c (x-y),
 \label{58}
\end{equation}
where
\begin{equation}
\Delta_c (x-y)=\theta\left(x_0
-y_0\right)\Delta_+(x-y)+\theta\left(y_0 -x_0\right)\Delta_-(x-y)
. \label{59}
\end{equation}
Using the equation \cite{Ahieser} $\left(\partial_\mu^2-\lambda
m\right)\Delta_c (x)=i\delta(x)$, one finds from Eq.(55),(58)
\begin{equation}
\left(\beta_\mu\partial_\mu+ m \overline{P}+\lambda P
\right)\langle T\Psi(x)\cdot\overline{\Psi}(y)\rangle_0 = i\frac{
m \overline{P}+\lambda P }{\lambda m}\beta_\mu\partial_\mu\delta
(x-y).
 \label{60}
\end{equation}
At $ A\neq 1$ ($m\neq\lambda$), we have non-local operators in
Eq.(58),(60) because the variable $p^2=-\partial_\mu^2$ enters the
$\lambda$. Expressions for PDK theory follow from Eq.(58),(60) at
$m=\lambda$ ($A=1$, $B=2$), and as a result, there are only local
operators.

\section{Discussion}

We have considered the generalized WTH equations, that describe
vector fields in the PDK formalism, Eq.(7). Eq.(7) is valid for
any parameters $A$ and $B$. For the Tucker-Hammer equation at
$A=1$, $B=2$ Eq.(7) reduces to PDK equation. For the Weinberg
equation, at $A=0$, $B=1$, we obtain from Eq.(7) the generalized
PDK equation which is more complicated because it contains the
second derivatives. To consider solutions to the wave equation (7)
for particles in external electromagnetic fields, one has to make
the replacement $\partial_\mu\rightarrow\partial_\mu-ieA_\mu$
($A_\mu$ is the vector-potential of an electromagnetic field).
Then solutions to the Weinberg equation will be different compared
to solutions of the PDK equation. Therefore, physical observable
also will be different. The equation for the massless particles,
Eq.(16), is also different (at $C\neq 0$, $A\neq 1$) comparing
with the Maxwell equations in the matrix form \cite{Moroz},
\cite{Kruglov2}.

Solutions of equations obtained in the form of matrix-dyads allow
us to make calculations of processes in the quantum theory of
vector particles, described by generalized WTH equations, in the
simple manner. From the generalized PDK Lagrangian, we have found
the conserved electric current and energy-momentum tensor in the
matrix PDK form that are convenient for analyzing. Quantization of
fields and the propagator obtained make it possible to use the
perturbation theory for different quantum calculations.

\section{Appendix}

From Eq.(9),(19), we obtain
\[
\Lambda_\pm^2=-\widehat{p}^2+\left(m+\lambda\right)\Lambda_\pm-\lambda
m,~~~~~~~~~~~~~~~~~~~~~~~~~~~~~~~~~~~~~~~~~(A1)
\]
\[
\Lambda_\pm^3-\left(m+\lambda\right)\Lambda_\pm^2+\lambda
m\Lambda_\pm=\mp
ip^2\widehat{p}-m\widehat{p}^2\overline{P}-\lambda
\widehat{p}^2P,~~~~~~~~~~~~~(A2)
\]
where
\[
\widehat{p}^2\overline{P}=\overline{P}\widehat{p}^2=p^2\varepsilon^{\mu,\mu}
-p_\mu p_\nu\varepsilon^{\nu,\mu},
~~~~~~~~\widehat{p}^2P=P\widehat{p}^2 =p_\mu
p_\nu\varepsilon^{[\lambda\mu],[\lambda\nu]}.
\]
Squaring Eq.(A1) and using the relation
$\widehat{p}^3=p^2\widehat{p}$, after some algebraic
manipulations, we find the minimal polynomial of the matrix
$\Lambda_\pm$
\[
\left(\Lambda_\pm-m\right)\left(\Lambda_\pm-\lambda\right)\left[
\left(\Lambda_\pm-m\right)\left(\Lambda_\pm-\lambda\right)+p^2\right]=0.~~~~~~~~~~~~~~~(A3)
\]
On-shell, one has to put $p^2=-\lambda m$ in Eq.(A3). Eq.(A3)
holds for any $m$ and $\lambda$. At a particular case, at
$m=\lambda$ the equation is simplified and becomes
\[
\left(\Lambda_\pm-m\right)\left[
\left(\Lambda_\pm-m\right)^2+p^2\right]=0.~~~~~~~~~~~~~~~~~~~~~~~~~~~~~~~~~~~(A4)
\]
It should be noted that if we put $m=\lambda$ in Eq.(A3), we get
Eq.(A4) multiplied by the factor $\left(\Lambda_\pm-m\right)$, but
the real minimal equation is (A4). For the massless case the
matrix (35) obeys the equation as follows:
\[
\Lambda_\pm \left(\Lambda_\pm-\kappa\right)\left[\Lambda_\pm
\left(\Lambda_\pm-\kappa\right)+p^2\right]=0,~~~~~~~~~~~~~~~~~~~~~~~~~~~~~~~(A5)
\]
and on-shell $p^2=0$.

\end{document}